\documentclass[usegraphicx]{mn2e}
\begin{document}

\title[Neutrinos]{Neutrinos as cluster dark matter}

\author[R.H. Sanders] {R.H.~Sanders\\Kapteyn Astronomical Institute,
P.O.~Box 800,  9700 AV Groningen, The Netherlands}

 \date{received: ; accepted: }

\maketitle
\begin{abstract}

The dynamical mass of clusters of galaxies, calculated in terms of
modified Newtonian dynamics, is a factor of two or three times 
smaller than the Newtonian dynamical mass but remains significantly larger 
than the
observed baryonic mass in the form of hot gas and stars in galaxies.
Here I consider further the suggestion that the undetected matter
might be in the form of cosmological neutrinos with mass on the order
of 2 eV.   If the neutrinos and baryons
have comparable velocity dispersions
and if the two components maintain their cosmological density ratio,
then the electron density in the cores
of clusters should be proportional to $T^{3/2}$, as appears to be
true in non-cooling flow clusters.  This is equivalent to the 
``entropy floor'' proposed to explain the steepness of the
observed luminosity-temperature relation, but here preheating of the
medium is not required.
Two fluid (neutrino-baryon) hydrostatic models of clusters, in
the context of MOND, 
reproduce the observed luminosity-temperature relation of clusters.  If
the $\beta$ law is imposed the gas density distribution, 
then the self-consistent models predict the general form of the
observed temperature profile in both cooling and non-cooling flow clusters.

\end{abstract}
\section{Introduction}

In the past 25 years, Milgrom's proposed alternative to dark matter,
the modified Newtonian dynamics or MOND \cite{m83}, has enjoyed 
considerable phenomenological success with respect to galaxy 
scaling relations and individual rotation
curves (see Sanders \& McGaugh
2002 for a review).  The range of this predictive power has
recently been extended from very low luminosity objects \cite{ms07}
to luminous early-type spirals \cite{sn07}.  However, for some
time it has been recognised that MOND does not fully account for
the mass discrepancy in rich clusters of galaxies 
\cite{tw88,geal92,rhs99,asq,rhs03,posi05}.  This is basically
because, with MOND, the departure from Newtonian dynamics appears
below a critical acceleration , $a_0$.  The value of $a_0$ determined
from galaxy rotation curves is about $10^{-8}$ cm/s$^2$.  In clusters
of galaxies, the observed acceleration (estimated from the 
hydrostatic gas equation) is typically greater than
$a_0$ in the central regions, so MOND cannot alleviate the observed
discrepancy there.  When the MOND version of the hydrostatic 
gas equation is applied to the observed temperature and density
distributions in X-ray emitting clusters of galaxies,
one finds that there remains a discrepancy roughly of a factor
of three between
the identifiable baryonic mass in gas and stars and the dynamical
mass.  That is to say,
MOND also requires the presence of undetected, or dark, matter in
clusters of galaxies, albeit significantly less than conventional
Newtonian dynamics.

X-ray and weak lensing observations of the famous 
``bullet cluster'' \cite{cleal06} are now presented as definitive
evidence for non-dissipative dark matter in clusters of
galaxies and, by extension, as evidence against MOND.  Had it been previously 
claimed that MOND fully accounts for the kinematic
and X-ray observations of clusters without dark matter, then 
this 
object would indeed have been quite problematic for the theory.
But, in fact, the bullet creates no {\it additional} difficulties
for MOND; the quantity of dark matter required is consistent with that
suggested by the previous analyses (Angus et al. 2007).
What the bullet cluster adds (and this is a significant addition)
is convincing evidence that the dark component cannot be dissipative 
like the extended X-ray emitting gas.

At first thought, it might appear negative for MOND
that additional dark matter is required in the cluster
environment.
But one could also consider this to be a bold prediction-- 
that there is more matter to be discovered in clusters.
For example, there are more than enough undetected
baryons to make up the missing dark component; they need only be
present in some non-dissipative form which is difficult to observe.
Moreover, it is now known that at least two of the three active neutrino types
have non-zero mass \cite{fuku}.  Primordial neutrinos are
present in the Universe with a number density comparable to that of
photons, so non-baryonic dark matter certainly exists.  
If the neutrino mass scale is as large as 2 eV, then neutrinos
would comprise a non-baryonic component of
rich clusters \cite{rhs03}, but because of phase space constraints
\cite{tg79} they could not accumulate in individual galaxies.
$\beta$ decay
experiments have restricted the mass of the electron neutrino
to be less than about 2.2 eV, so it remains possible that
this is the appropriate mass scale \cite{groom}.
The addition of cosmological neutrinos with mass of this order
produces a reasonable fit to  
the observed angular power spectrum of the cosmic microwave
background radiation, through the second peak, without the presence of cold
dark matter \cite{ssmcmb}.

Therefore, here I consider further the possibility 
that the missing component in
clusters of galaxies in the context of MOND may be neutrinos.
I demonstrate that the form of correlations between observable quantities--
the size-temperature relation ($R\propto T$)
and the gas mass-temperature relation ($M_g\propto T^2$)
are consistent with MOND expectations without any additional astrophysical
mechanism other than violent relaxation.  
However, the luminosity-temperature relation ($L\propto T^3$)
is not
consistent with MOND by itself; an additional dark component 
seems to be required \cite{rhs03}.  

For a sample of well-studied clusters, the MOND dynamical mass is proportional
to but a factor of three or four times larger than the observable gas mass. 
This suggests that the detectable 
baryonic mass is a fixed fraction of the dark mass.  If the
mass of each active neutrino type is about 2 eV, then
the ratio of neutrino to baryonic cosmological mass densities would be 
about 2.8, consistent with the inferred (via MOND) ratio of gas to dark matter
in clusters.

I consider the structure of two-component isothermal objects consisting
of neutrinos and gas.  The neutrino fluid is described by the equation
of state of partially degenerate fermions, where the degree of degeneracy
in the centre is arbitrarily set to a specified value. 
Because the 
maximum density of neutrinos is proportional to $T^{3/2}$ this means
that, in a uniformly mixed two-component fluid, the central
gas density would also be proportional to $T^{3/2}$.  This appears to
be the case in non-cooling flow clusters where there has been no
inward flow of gas resulting in a rearrangement of the cosmological density
ratio of baryons-to-neutrinos.  With such models, the observed luminosity-
temperature scaling relation for clusters is recovered primarily because
$L\propto {n_e}^2 \propto T^3$.  

I then consider individual clusters with the two-fluid model.  After
specifying the mass scale of the neutrinos and the degree of
degeneracy in the centre, I assume that the
presumably constant neutrino velocity dispersion is equal to
that of the gas as implied by the mean emission-weighted gas temperature. 
Given the observed gas density distribution
(via $\beta$-model fits to the X-ray intensity distribution) I calculate
the radial dependence of the gas temperature 
which is consistent with this density
distribution.  The only free parameter is the central gas temperature
which must lie in a narrow range in order to yield a sensible
temperature distribution (one in which the temperature does not
rapidly increase to large values or fall to zero).  These calculated
temperature distributions are found to be generally consistent with
those now observed-- a slowly decreasing temperature for non-cooling flow
clusters, and a temperature which first increases with radius and
then decreases for cooling flow clusters.

I conclude that the two-fluid neutrino-baryon model for clusters of
galaxies in the context of MOND, is consistent both with the
correlations between observable quantities and with the observed density
and temperature distributions in individual clusters.

In all that follows I have scaled observational results and
correlations to $H_0 = 72$ km/s-Mpc. 

\section{Global scaling relations for clusters of galaxies}

It is of interest to consider correlations between the observed, and
not the inferred, properties of X-ray emitting clusters of galaxies.
One of the most obvious correlations is that between the cluster
size (out to a specified X-ray intensity level) and the temperature
of the hot gas.  Mohr et al. (2000) find that this relation,
for nearby clusters, is:
$$R= 0.5 \Bigl({T\over{6\, keV}}\Bigr)^{1.02}\,\, {\rm Mpc}\eqno(1)$$ 
with a surprisingly low
scatter of about 15\%.  The expectation for self-similar collapse
is more like $R\propto T^\alpha$ where $\alpha \approx 1/2 - 2/3$.
Mohr et al. explain the difference between the observations and the
expectations by noting that the fraction of gas (as opposed to the total
baryonic fraction including the luminous matter in galaxies) appears
to increase with cluster temperature.  While this seems
to be a general trend \cite{deal90,es91}, the observed
relationship between gas mass/total baryonic mass
and temperature exhibits enormous scatter.  It is unclear
how such a tight correlation can survive.

In terms of MOND, eq.\ 1 takes on a different meaning.  The mean
internal gravitational acceleration in clusters is given roughly by
$a = \sigma^2/R$ ($\sigma$ is the velocity dispersion and proportional
to temperature).  Then we may interpret eq.\ 1 as defining a
single characteristic internal acceleration which is independent
of cluster size or temperature; i.e., $a \propto T/R \approx 6\times 10^{-9}$
cm/s$^2$.  This is within a factor of two of the MOND critical
acceleration ($a_0\approx 10^{-8}$ cm/s$^2$).  It has been 
pointed out previously that in pressure supported, nearly isothermal systems,
the internal acceleration is approximately the MOND acceleration \cite{sm02}.
Viewed in this way, the cluster temperature-size relation simply
reflects that characteristic internal acceleration 
and requires no astrophysical input other than the requirement of
a near isothermal state, presumably due to violent relaxation.

A second correlation is that between the temperature
and the observed gas mass.  Mohr et al. (1999) give
$${{M_g}\over{10^{14}M_\odot}}\approx 0.017\, {T_{keV}}^2.\eqno(2)$$
For self-similar collapse, making use of the Newtonian
virial theorem, the expectation would be
$M_g\propto T^{3/2}$.  The disagreement between observations and
expectations is again attributed to a systematically changing fraction
of baryons in hot gas perhaps resulting from increasing energy
injection into the intra-cluster medium  or
a decreased efficiency of galaxy formation in hotter clusters. 

With MOND, however, eq.\ 2 would be the expectation from the dynamics
of isothermal (or semi-isothermal) spheres;
it is, in effect, an extrapolation of the Faber-Jackson relation
for elliptical galaxies \cite{rhs94}
and, in the case of MOND would apply to all pressure supported,
near isothermal 
systems \cite{m84}.  Basically, the equation of
hydrostatic equilibrium in the MOND regime yields 
$${M\over {10^{14}M_\odot}} \approx .06\, {T_{kev}}^2 \eqno(3)$$ for the total
dynamical mass in terms of MOND.  Here I have assumed an isothermal
$\beta$-model with an outer logarithmic density gradient of
$-3\beta$ where $\beta = 0.6$ on average.  Although the form
of the temperature-gas mass relation is predicted, the 
normalisation is not; the MOND dynamical mass implied by eq.\ 3
is roughly 3.6 times larger than the observed gas mass.

The third correlation, and the most discussed in the literature, is
that between luminosity and temperature.  Again the correlation
is well-fitted by a power-law; for example, Ikebe et al. (2002) give
$$L_x = {4\times 10^{42}} {T_{keV}}^{2.5} {\rm ergs/s^2}\eqno(4)$$
for the total X-ray luminosity between energies of 0.1 to 2.4 keV.
Others find exponents closer to 3 with suggestion of a steepening
for cooler clusters \cite{ae99}.  The expectation for 
self similar cluster formation in the context of CDM
is a shallower power law $L\propto T^2$.  Here, preheating of
the intergalactic medium (heating before cluster formation)
by winds from forming galaxies
is invoked to to provide an ``entropy floor'' for the 
subsequent intra-cluster medium (entropy is defined as
$T_{keV}/{n_e}^{2/3}$).
This would have the effect of inflating the
the intra-cluster medium particularly in lower temperature galaxies and 
steepening the predicted luminosity-temperature relation
\cite{pcn99}.  

MOND, in this case, fairs worse.  In MOND a virialized
system has a characteristic density $\rho_M \propto
{a_0}^2/(GT)$.  As we saw above, the characteristic radius of
the cluster is $R\propto T/a_0$.  For free-free
radiation, $L\propto \rho^2 T^{1/2} R^3$; therefore, the naive
prediction would be $L\propto T^{3/2}$.  I discussed this point
previously (Sanders 2003) and noted that a dark matter
component with a constant density and a core radius roughly twice
that of the gas core radius could bring the MOND prediction
in line with the observations.

\section{The magnitude of the discrepancy with MOND}

MOND predicts more mass than is directly
observed in clusters of galaxies.  Determination of the magnitude
of this discrepancy, however, is not straightforward.
For an isothermal gas, the  
hydrostatic gas equation in the MOND limit implies that the
mass within radius R is given by
$$M_d = \alpha^2\Bigl({{kT}\over {wm_p}}\Bigr)^2 (G a_0)^{-1} \eqno(5)$$
where $w=0.62$ is the mean atomic weight, $m_p$ is the proton mass, and
$\alpha=d\,ln\,\rho/d\,ln\,r$ is the logarithmic density
gradient evaluated at R.  The gas density distribution is typically
described by the $\beta$ model used to fit the X-ray surface brightness
distribution:
$$\rho= \rho_0\Bigl[1+\bigl({r\over {r_c}}\bigr)^2\Bigr]^{-1.5\beta}.\eqno(6)$$
Therefore beyond three or four core radii $\alpha\rightarrow -3\beta$--
i.e., the predicted dynamical mass converges.  On the other hand,
$\beta= 0.5-1.0$ which means that the gas mass 
continues to increase with radius
(the $\beta$ model surely must have a limited range of
viability).  Therefore, the magnitude of the discrepancy
(defined as $M_d/M_g$), as estimated from the $\beta$ model,
decreases with radius; i.e., the discrepancy depends
upon the radius within which the estimate is made.

I have determined the MOND dynamical mass and gas mass 
for a well-studied sample of clusters given by Reiprech \& B\"ohringer (2002).
I make use of the radius-temperature relation to estimate
the cutoff radius, but I take this limiting radius to
be about 20\% larger than that given by eq.\ 1.  This implies
that the cutoff 
is, on average, about six core radii which is a probable
range of validity for the $\beta$ model fit.  Since the accelerations
in the central regions of the clusters are typically on the order
of $a_0$, the form of the MOND interpolating function plays a role;
here I take the simple form suggested by Zhao and Famaey (2005):
$$\mu(x) = {{x}\over{1+x}} \eqno(7)$$
where $x=a/a_o$ with $a=\alpha kT/(wm_pR)$.  The MOND dynamical mass
is then given by 
$$M_m={{xM_N}\over{1+x}}\eqno(8)$$
where the Newtonian mass is $M_N = -\alpha RkT/(Gwm_p)$.
The gas is only a fraction of the total baryonic mass-- the remainder
being in the form of stars in galaxies.  As before, I take the ratio
of intra-cluster gas to stellar mass as given by $T_{keV}/2.5$ \cite{rhs99}.
There is considerable scatter about this approximate relation.

\begin{figure}
\resizebox{\hsize}{!}{\includegraphics{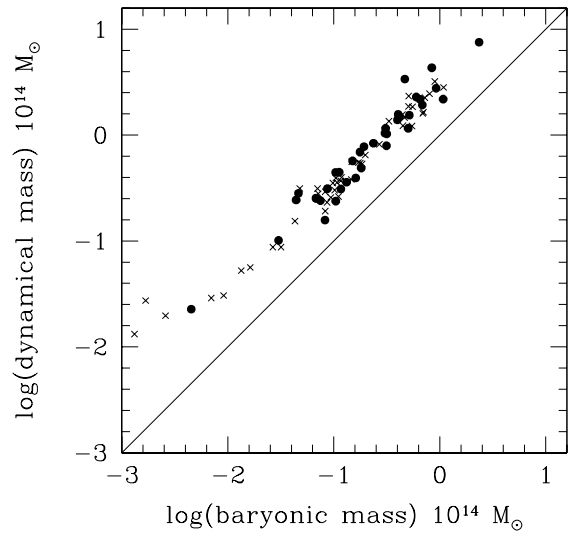}}
\caption[]{The logarithm of the MOND dynamical mass plotted against
the logarithm of inferred baryonic mass (both in units of $10^{14}M_\odot$).
The crosses indicate cooling flow clusters (those where
the central cooling timescale is less than the Hubble time), 
and the solid points are non-cooling flow clusters.}
\end{figure}

The result of this analysis is shown in Fig.\ 1 which is a plot
of the MOND dynamical mass against the inferred baryonic mass
out to a limiting radius given by $100\,T_{keV}$ kpc.  The ratio of
the MOND dynamical mass to the baryonic mass is, on average,
3.7 with no significant difference between the cooling flow and
non-cooling flow galaxies.  That is to say, the observed baryonic mass in
clusters accounts for one-third to one-forth of the MOND dynamical mass.

It is evident from Fig.\ 1 that the ratio of dynamical to baryonic
mass is roughly independent of the baryonic mass.  
There is an indication of a
deviation at low values of the baryonic mass possibly because of
an increased fraction of matter in the stellar component of galaxies,
but overall, the points lie parallel to the line of equality. 
This is of interest because it
suggests that a fixed cosmic ratio of dark to baryonic
matter may be sampled by the cluster medium.

\section{The role of massive cosmological neutrinos}

If neutrinos are more massive than a few tenths of an electron
volt, then, because of the small mass differences, all 
types have about the same mass.  The cosmological 
density of neutrinos would then be given by
$\Omega_\nu = 0.062 m_\nu$
where $m_\nu$ is the mass of a single neutrino type in eV.
In this relation I have assumed three left-handed neutrino types and
their anti-particles, but
the formula would apply for Majorana as well as Dirac neutrinos.

Given that the cosmological density of baryons is $\Omega_b = 0.044$
(e.g. Spergel et al. 2006, with h=0.72),
this means that $$\Omega_\nu/\Omega_b = 1.4\, m_\nu. \eqno(9)$$  
If we suppose that clusters of galaxies sample this universal neutrino-
baryon density ratio then the dynamical mass-to-baryonic mass ratio
implied by Fig.\ 1 would mean that $m_\nu \approx 2$ eV.

Cosmological neutrinos freeze out of the primordial fireball
below temperatures characteristic of the weak interaction
scale of a few meV.  The phase space distribution is Fermi-Dirac
with a maximum for each species of
one particle per cell with volume $2h^3$; i.e., one-half that of
complete degeneracy.  This initial phase space density is maintained
as a limit on the final phase space density of any collapsed virialized
object (Tremaine \& Gunn 1979); of course, 
the degeneracy limit is absolute.  

In the formation of a cluster scale object out of
the mixed fluid of neutrinos and baryons, it is expected that
the two fluids attain the same velocity dispersion via
violent relaxation \cite{ktb97}.  With this assumption, the maximum
contribution of each neutrino type (including its anti-particle) 
to the density of the neutrino fluid is
$$\rho_\nu = (2\pi)^{3\over 2} {m_\nu}^4 \sigma^3 h^{-3}\eqno(10)$$
\cite{ktb96}.  For three neutrino types (and their anti-particles)
this translates to a mass density of
$$\rho_{max} = {1\times 10^{-28}}{\Bigl[{{m_\nu}\over {1\, {\rm eV}}}\Bigr]}^4
 {T_{keV}}^{3/2} {\rm gcm^{-3}}
\eqno(11)$$
If the ratio of the
densities of baryons and neutrinos in clusters is the same
as the cosmological ratio (eq.\ 9), this means that the density of 
electrons, in a fully ionised plasma, would be 
$$n_e = 3.5\times 10^{-5}{\Bigl[{{m_\nu}\over {1\, {\rm eV}}}\Bigr]}^3
 {T_{keV}}^{3/2} {\rm cm^{-3}}. \eqno(12)$$

Therefore, if the ratio of baryon to neutrino densities in clusters
is identical to the cosmic ratio, as one might expect if there has
been no subsequent cooling and inflow of baryons, then the central
electron density should increase as $T^{3/2}$.  This is an observational
prediction which can be tested, and in Fig.\ 2 we see the central
electron density of in the cluster sample of Reiprech \& B\"ohringer
plotted against the mean electron temperature.  The solid points
indicate those clusters where the cooling time is more than
$10^{10}$ years (non-cooling flow clusters) and the crosses are those
clusters with cooling timescales less than this value (cooling flow
clusters).  The solid line is eq.\ 12 with $m_\nu = 1.9$ eV.
We see that in those objects where no cooling and inflow
of gas is expected-- where the cosmic ratio of baryons to neutrinos
is maintained-- there does appear to be such a correlation.

\begin{figure}
\resizebox{\hsize}{!}{\includegraphics{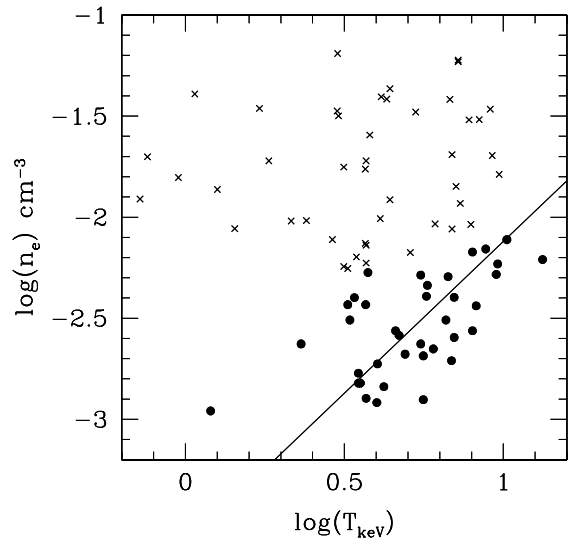}}
\caption[]{The logarithm of of the central electron
density (from the fitted $\beta$ models of Reiprech \&
B\"ohringer) vs. the logarithm of the temperature in eV.
The points are the non-cooling flow clusters and the crosses
are the cooling flow clusters.  The line corresponds
to eq.\ 12 for a neutrino mass of 1.9 eV.}
\end{figure}

It is interesting to note that the electron density-temperature relation
imposed by the neutrino density limit and the assumption of fixed
baryon-to-neutrino ratio (eq. 12) would also correspond to a 
constant entropy, $T_{keV}/{n_e}^{2/3}$.  Imposing such a temperature 
dependent
limit on the electron density is equivalent to imposing
an entropy floor.  {\it This
is, in fact, the very mechanism which has been suggested as to steepen
the predicted luminosity-temperature relation, but here the limit
is not generated by preheating but by the phase space constraints
on the neutrino fluid.}  It is interesting that the entropy limit
corresponding to eq.\ 12 would be 230 keV cm$^2$ for $m_\nu=2$ eV
which is
very near the value proposed to solve problem of cluster scaling
relations \cite{ldpc}.

Before considering the implications for the
luminosity-temperature relation it is necessary to
describe, in a general way, the structure of a self-gravitating 
object consisting entirely of neutrinos near the degeneracy limit.
The object can be approximated as a $\gamma=5/3$ polytrope;
the structure would be roughly that of a constant density core with
a rapid decline beyond a core radius.  The core radius can be
estimated by combining the density temperature relation (eq.\
10 or 11) with an appropriate temperature-mass (or viral relation).
There are two such possible relations.  The first, for Newtonian dynamics, is
$$\sigma^2 = {{4\pi}\over 10.5}G\rho_c {R_c}^2 \eqno(13)$$
where $\sigma$ is the one-dimensional velocity dispersion, and
I have taken the potential energy of a $\gamma=5/3$ polytropic sphere.
The second is for modified Newtonian dynamics,
$$\sigma^4 = {16\pi\over 243} G\rho_c {R_c}^3 a_0 \eqno(14)$$
\cite{m98}.
Then with eq.\ 11 we find two expressions for the core radius:
Newtonian,
$$R_c = 4.5 {\Bigl[{{m_\nu}\over {1\, {\rm eV}}}\Bigr]}^{-2} 
{T_{keV}}^{-{1\over 4}} {\rm Mpc}\eqno(15)$$
and MOND,
$${R_c}^* = 1.8 {\Bigl[{{m_\nu}\over {1\, {\rm eV}}}\Bigr]}^{-{4\over 3}} 
{T_{keV}}^{1\over 6}{\rm Mpc}.
\eqno(16)$$
These two values are equal when 
$$T_{keV} = 9{\Bigl[{{m_\nu}\over {1\, {\rm eV}}}\Bigr]}^{-{8\over 5}}.
\eqno(17)$$
This meaning of the final expression is that when the temperature is 
less than about 3 keV (for $m_\nu\approx 2$ eV) the cluster core is in
the MOND regime and eq.\ 16 applies; 
for higher temperatures the cluster core is the Newtonian regime with 
core radius given by eq.\ 15.  In either case,
this characteristic 
radius is on the order of 800 kpc and depends very  weakly on
temperature in both regimes.

Now assuming that baryons are mixed into this cluster with the
cosmic ratio, most of the X-ray emission would be coming from
the core region.  The X-ray luminosity is
$L\propto {n_e}^2 T^{1/2} {R_c}^3$ or
$$L\propto T^4$$ for cooler clusters (MOND regime) and 
$$L\propto T^3$$ for warmer clusters (Newtonian regime).
In this way, the steeper than expected luminosity-temperature relation
can be understood in terms of the mixed neutrino-baryon fluid.  The
transition from Newton to MOND would explain the steepening of
the correlation for lower temperature clusters.

This, of course, is only approximate.  X-ray emission also originates
beyond the core, and the gas is not isothermal.
In the next section I derive detailed two fluid models of clusters.

\section{MOND-neutrino-baryon models of clusters} 

The contribution of neutrinos to the mass budget of clusters has
been considered previously in the context of mixed CDM-HDM 
models \cite{keal,tkb,nakmor}.  In the process of collapse and 
virialization of
a multi-fluid mixture of neutrinos and baryons (and/or CDM), 
violent relaxation \cite{dlb67} is expected to produce
equal velocity dispersions for the various components, i.e.,
$\sigma_b=\sigma_\nu$.  I assume this condition here, although
it may not be strictly true; for example, violent relaxation may be incomplete
in the outer regions of clusters. Moreover, recent observations 
have demonstrated that the hot gas in clusters is not isothermal, 
in general.  The implications for the hypothetical neutrino 
fluid are unclear,
but I will assume an isothermal condition with the neutrino velocity 
dispersion equal the emission-weighted velocity dispersion of the gas.

It should also be kept in mind that the density given by
eq.\ 11 is an upper limit to the density of the neutrino fluid;
this value follows from the limit imposed by
initial phase space density via the collision-less Boltzmann equation
\cite{tg79};  the course grained phase space density in
a final relaxed virialized system, and hence the space density, may be less
significantly less than this value.
Moreover, while the neutrino fluid may be partially degenerate
in the core, it will certainly depart from degeneracy in the outer
regions where the space density declines and the velocity dispersion
remains constant.  Therefore, with respect to the neutrino fluid,
we need to consider the equation of state of partially degenerate
matter.

With these caveats in mind, I calculate the structure of the
neutrino-baryon fluid by applying the hydrostatic gas equation
separately to each fluid:
$$ {1\over\rho_i} {{dP_i}\over {dr}} = -g \eqno(18)$$
where the subscript $i$ refers to one of the two fluid components,
$P_i$ is the pressure of that component, $\rho_i$ is the density, 
and $g$ is the total gravitational acceleration resulting from the
two components and given by the simple MOND expression,
$$g\mu(g/a_0) = g_N\eqno(19)$$
with $\mu$ given by eq.\ 7 and the $g_N$, the Newtonian force,
given by
$ g_N = G(M_\nu(r)+M_b(r))/r^2$

The pressure of the baryonic component
is $$P_b=\rho_b{\sigma_b}^2 \eqno(20)$$ 
The velocity dispersion for the baryons ${\sigma_b}^2$ (the
temperature) may be specified as a constant (isothermal) or may 
vary with radius if the density distribution is specified.

The equation of state for a partially degenerate neutrino gas is
given parametrically by
$$\rho_\nu = {{8\pi g}\over\sqrt{2}}\eta {\sigma_\nu}^3 F_{1\over 2}(\chi)
\eqno(21)$$ and
$$P_\nu = {{8\pi\sqrt{2} g}\over 3} \eta {\sigma_\nu}^5 F_{3\over 2}(\chi)
\eqno(22)$$
where $g$ is the statistical weight, $\eta={m_\nu}^4/h^3$, and
$$F_p(\chi) = \int_0^\infty {x^p[1+exp(x-\chi)]^{-1} dx} \eqno(23)$$   
(see e.g. Landau \& Lifshitz 1980).  Here the degeneracy factor $\chi$
is the chemical potential divided by the temperature; large
positive values of $\chi$ would correspond to complete degeneracy 
($P_\nu \propto {\rho_\nu}^{5/3}$) and
large negative values, to complete non-degeneracy ($P_\nu\propto \rho_\nu
{\sigma_\nu}^2$).  

The density and the velocity dispersion formally
determine the chemical potential, but there is no independent
method, short of numerical calculations of collapse of the two-
fluid mixture to estimate $\chi$. This essentially depends upon
how close the actual central density of the neutrino fluid is to
the maximum permitted by the phase space constraints and the
effectiveness of violent relaxation in equalising the velocity
dispersions of the two fluids.  There are
various arguments suggesting that the true density should range
between 10\% and 100\% $\rho_{max}$ \cite{madep,keal,nakmor}; 
here I take $\rho_\nu\approx
\rho_{max}$ because this is in rough agreement with the observational
results shown in Fig.\ 2 (i.e., the central neutrino density is 
50\% that corresponding to quantum mechanical degeneracy).  This
corresponds to $\chi = -0.5$, and in the calculations described below, I 
have fixed the central value of $\chi$ at this value.  
Realistically, due to random initial conditions for cluster formation, 
one would expect a dispersion in this parameter.

\begin{figure}
\resizebox{\hsize}{!}{\includegraphics{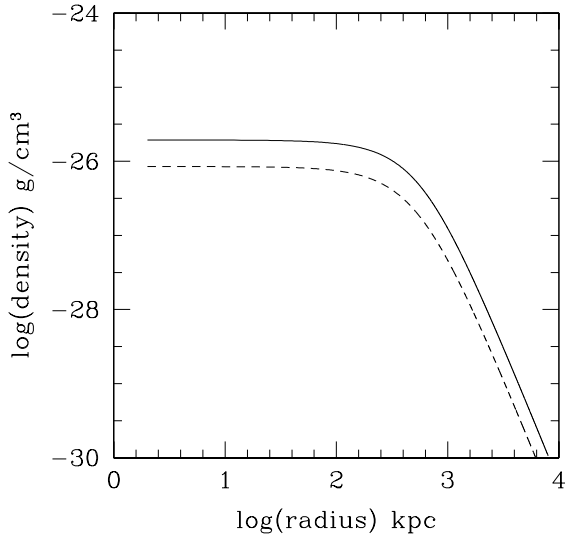}}
\caption[]{The logarithm of of the neutrino fluid density
with $m_\nu = 1.9$ eV
(solid curve) and the baryon fluid density (dashed curve)
as a function of the logarithm of radius. The two
fluids are assumed to be isothermal corresponding to the gas temperature of
5 keV and to have a mass ratio equal to the cosmic density ratio (2.68 in
this case).
The central degeneracy factor is $\chi=-0.5$ implying
that the central neutrino density is about 0.4 of the
maximum density imposed by the initial phase space 
constraint (eq.\ 11). 
The accelerations near the core radius ($\approx 200$) kpc
are comparable to $a_0$}
\end{figure}

Assuming a comparable mass ($m_\nu=1.9$ eV) for all three neutrino types, 
a constant temperature for both fluids (with the velocity dispersion
of neutrinos equal to that of the baryons), and a mass ratio of the
two components equal to the cosmic density ratio (2.68 in this case),
the structure of the object may be determined by numerical integration of  
eqs.\ 18 supplemented by eqs.\ 19-23.  With these assumptions and 
constraints, the structure completely determined
when the gas temperature is specified.  Fig.\ 3 shows the density distribution
for the two fluids in the case where $T_{keV}= 6$.  The density
of the two components effectively track each other in the inner regions.
The distributions can be described as a roughly constant density core
extending to about 300 kpc, followed by a rapid decrease (asymptotically
$\rho \propto r^{-3.5}$ for both components.  

The X-ray luminosity resulting from free-free emission of
the hot ionised gas may also be calculated for any such object.  The
resulting luminosity-temperature relation is shown in Fig.\ 4 compared
to the sample of Ikebe et al. (2002).  This is similar to the figure
shown in Sanders (2003) but there the dark component was added arbitrarily,
with no underlying physics.  Here the X-ray emission is that from 
self-consistent neutrino-baryon fluid spheres with only the
neutrino mass and degeneracy factor specified arbitrarily.
The results generally agree with the observed relation; in particular
the steepening of the relation for low temperature clusters is evident.

\begin{figure}
\resizebox{\hsize}{!}{\includegraphics{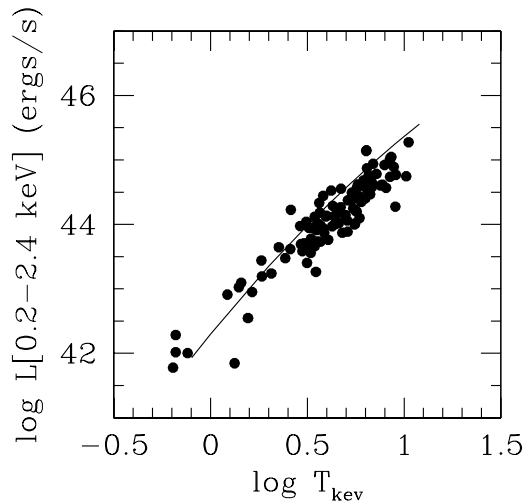}}
\caption[]{A log-log plot of the luminosity-temperature
relation for self consistent MOND-neutrino-baryon fluid spheres 
compared to the observations of Ikebe et al. (2002).  The neutrino
mass is taken to be 1.9 eV and the degeneracy factor is -0.5.
With the assumptions of isothermal state, equal velocity dispersions,
and cosmic ratio of baryons-to neutrinos the structure and X-ray
luminosity of these objects is completely specified.}
\end{figure}

\section{Non-isothermal $\beta$ models}

The gas density distribution in real clusters does not resemble
that shown in Fig.\ 3.  The density in these isothermal
models falls too rapidly beyond the
core to be consistent with the observed distribution of X-ray
surface brightness in clusters; i.e., the power law implied by
the fitted $\beta$ models (eq.\ 6) is more like $r^{-2}$ rather
than $r^{-3.5}$.  However, it is now known that the
gas in clusters cannot be described as isothermal.  There
appears to be a characteristic temperature profile in
clusters: in non-cooling flow clusters
the temperature declines by 20 or 30\%
between 100 kpc and 1 Mpc; in
cooling flow clusters there is first a rapid rise in the inner core region
(by as much as a factor of two) followed by the decline observed
in non-cooling flow clusters \cite {dgm,pjkt,preal,vikeal}

Therefore, here I take a different approach.  I assume the density
distribution of the gas is given by a $\beta$ model and I solve
eq.\ 19 to determine the temperature distribution for the gas.
The neutrino fluid is again assumed to be isothermal with a 
velocity dispersion equal to that implied by mean emission weighted 
temperature of the gas.  
In this way, one may predict the temperature distribution
in a particular cluster with a fitted $\beta$ model.  The only
adjustable parameter is the central value of the gas temperature,
and this is chosen to match either the mean emission-weighted
temperature of the gas or to achieve the best fit to the
projected gas temperature as a function of radius.  In fact,
there is generally a narrow range of this parameter which yields
a reasonable temperature distribution for the cluster-- one which
does not increase rapidly to absurdly high values or one which 
does not fall rapidly to zero.  

\begin{figure}
\resizebox{\hsize}{!}{\includegraphics{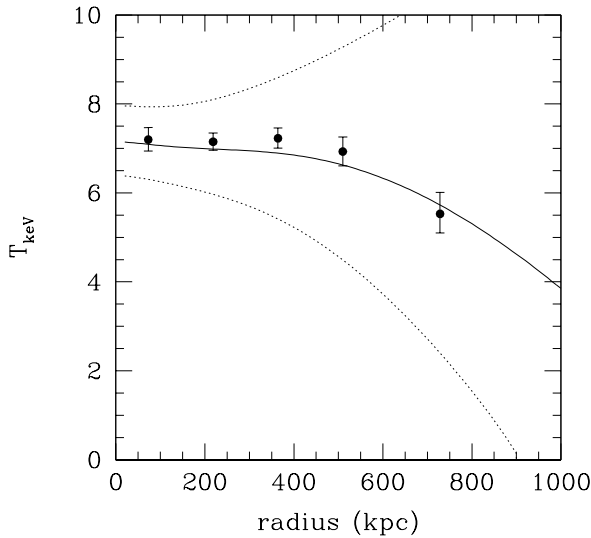}}
\caption[]{The predicted projected temperature distribution in
A2256 compared to that observed (de Grandi \& Molendi).
The assumed neutrino mass is 1.9 eV.  The $\beta$ model parameters
are 
the mean emission weighted gas temperature is 7.2 keV.  The neutrinos
are assumed to have a constant velocity dispersion appropriate to
this temperature.  The solid curve shows the prediction
when the central gas temperature is take to be 7.5 keV and the
dashed curves show the distributions when the central temperature
is taken to be 8 keV (upper curve) or 7 keV (lower curve).}
\end{figure}

This is illustrated in Fig.\ 5 which is the radial
dependence of projected gas temperature for the 
model fit to A2256 with $m_\nu= 1.9$ eV and $\chi = -0.5$.  
This is a non-cooling flow cluster
with a low central density ($n_e = .0031$ cm$^-3$) and a large core
radius ($r_c = 419$ kpc).  The decline of surface brightness
implies $\beta = 0.914$ \cite{rb02}.  Taking a central gas temperature of
7.5 keV produces a projected temperature profile (solid curve)
which is consistent with that observed \cite{dgm}.  The dashed 
curves illustrate the effect of taking a central gas temperature
one-half keV higher or lower; it is evident that the resulting
temperature profile is strongly dependent upon the initial
assumed central gas temperature.  This procedure has been carried
out for several non-cooling flow clusters, and the characteristic
temperature profile is as shown here; i.e., it is consistent
with the observed profiles for non-cooling flow clusters.

For cooling flow clusters, the results are even more sensitively
dependent upon the assumed central temperature, but, in
general, the calculated temperature profiles agree with those
observed for cooling flow clusters.  This is shown in Fig.\ 6
which illustrates the predicted temperature profile (again with
$m_\nu = 1.9$ eV and $\chi=-0.5$)
for the cooling flow cluster A85 with $\beta$ model parameters of
$\beta = 0.532$, $r_c=58.1$ kpc, $n_e=0.0204$ cm$^{-3}$ \cite{rb02}.  This
is compared to the observed temperature profile by De Grandi
\& Molendi.  The dotted curves show the effect of increasing
or decreasing the assumed central gas temperature by 0.1 keV;
i.e., the results here are extremely dependent upon the
this parameter.

For clusters with $\beta$ model fits characterised by
a small core radius and high central electron density, this 
is the general pattern predicted the two fluid models:  
a rapid rise in temperature followed by a gradual decline.  In general,
the predicted central temperature is lower than observed and the detailed
agreement is less impressive than for non-cooling flow clusters.
It should be kept in mind, however, that no baryonic component
other than the gas is included in these calculations; 
the galaxies, and in particular, large central cD galaxies are not
part of the mass modelling.

Overall, it appears that when the $\beta$ model is imposed upon
the gas density distribution, the temperature distribution required by
self-consistency is in general agreement with the observed
temperature profiles.
 
\begin{figure}
\resizebox{\hsize}{!}{\includegraphics{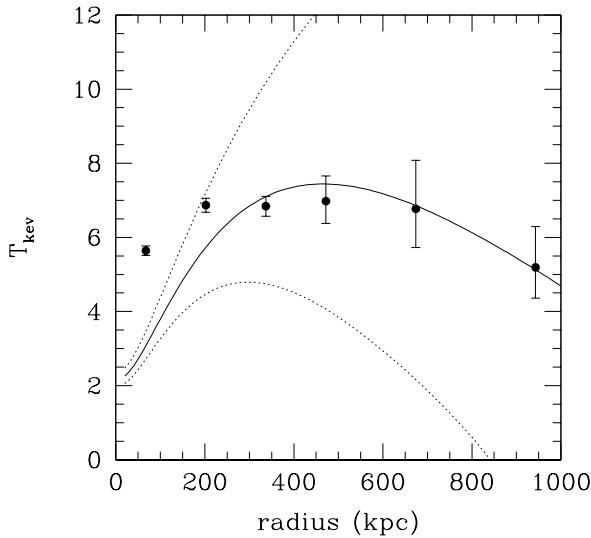}}
\caption[]{The predicted projected temperature distribution in
A85 (solid curve) compared to that observed (de Grandi \& Molendi).
The assumed neutrino mass is 1.9 eV and the degeneracy factor
is $\chi=-0.5$.  The $\beta$ model parameters
are $\beta = 0.532$, $r_c = 58.1$ kpc, $n_e = .0204$ cm$^-3$.
The central gas temperature is set to 1.69 keV.
The two dotted curves are the predicted temperature distributions
when the central gas temperature is set to 1.79 keV (upper curve)
and 1.59 keV.  This illustrates the extreme sensitivity of the
predicted distribution to central gas temperature in cooling flow
clusters (high central electron density, small core radius).}
\end{figure}

\section{Can individual galaxies have neutrino dark matter halos?}

It would 
seem possible that individual galaxies could possess extensive neutrino 
dark halos, even if the neutrino mass is as low as 2 eV.  
It is true that no primordial fluctuations on the
scale of galaxies would survive in the cosmic neutrino fluid due to
free streaming.  However, in the context of MOND, the baryonic component
would collapse first and act as a seed for subsequent neutrino
infall.  

The neutrino halo could have a mass almost three times larger
than the baryonic galaxy, but because of the phase space constraint,
it would be very extensive.  Rewriting eq.\ 16 in terms of rotation
velocity, $V_r$, we would find, 
$${R_c}^* = 1.3 {\Bigl[{{m_\nu}\over {1\, {\rm eV}}}\Bigr]}^{-{4\over 3}} 
\Bigl[{{V_r}\over {200\,{\rm km/s}}}\Bigr]^{1\over 3}
 {\rm Mpc}\eqno(24)$$
for $m_\nu=2$ eV the halo would extend to 500 kpc for a typical massive 
galaxy.  The total mass of the neutrino halo on galactic scale 
(10-20 kpc) would be less than 1\% of the galaxy's baryonic mass and
the contribution of the halo to the observed rotation curve in 
the outer regions would be less than 10 km/s.  Therefore, such
a halo could have no effect on observed galaxy kinematics.

It would, however, affect the weak lensing properties of galaxies
on large scale.  On the scale of hundreds of
kpc, the halo mass would exceed the mass of the galaxy, and 
dominate the lensing signal.  If, for example, the halo were 
flattened, then, with sufficient statistics, 
this could be apparent from the pattern of
background galaxy images \cite{hoekeal}.

\section{Conclusions}

Non-baryonic dark matter in the form of active neutrinos
certainly exists; only its contribution to the total mean
cosmic density is in question.
If the mass of the three neutrino types is as large as 2 eV,
then neutrinos will be a constituent of massive rich clusters
of galaxies.  For this mass scale, the cosmological
density ratio is of the order of the remaining mass discrepancy in clusters
calculated via MOND; i.e., such neutrinos would complete the
mass budget of clusters in the context of MOND.  Maintaining the
cosmological mass ratio in clusters leads naturally to a dependence
of central electron density on temperature ($n_e\propto T^{3/2}$)--
a dependence which is consistent with density inferred from
$\beta$ model fits to the
X-ray intensity distribution in non-cooling flow clusters
(Fig.\ 2).  
This gas density-temperature dependence corresponds precisely
to the constant entropy floor that has been proposed to account
for the steepness of the observed luminosity-temperature relation,
but in this case no preheating of the cosmic baryon fluid is
needed; the floor is provided by the phase space constraints
on neutrino density.  Of course, in cooling flow clusters one would expect
there to be an inward flow of gas and breaking of the cosmological
density ratio-- with resulting higher electron densities.  

Assuming that the
cosmic density ratio of the two fluids is maintained and that
the neutrinos and baryons have the same velocity dispersion, then
only the neutrino mass and the
degeneracy factor need be specified in order to calculate hydrostatic
two fluid models of clusters.  Such models, with $m_\nu \approx 2$ eV
exhibit the observed X-ray luminosity-temperature relation--
including a break to a steeper relation
for lower temperature clusters (Fig.\ 4).  If the gas density distribution
is constrained to follow the fitted $\beta$ models for clusters,
then the temperature distribution of the gas required for
self-consistency resembles that in actual clusters-- a general
decline over hundreds of kpc for non-cooling clusters, but a rapid
rise followed by a decline for cooling flow clusters (Figs. 5 \& 6).

Thus these calculations support the suggestion that the missing
component in clusters of galaxies, in the context of MOND, may
be neutrinos with mass near the present experimental upper limit.
A remaining observational problem with this is that some
clusters-- primarily cooling flow clusters-- apparently require
a higher central density of dark material than is permitted by
the phase space constraints on neutrinos \cite{rhs03}.  This is
evident both from X-ray observations and observations of
strong gravitational lensing in the centres of some clusters
where the implied central mass within an Einstein ring radius
(100-200 kpc) may be in excess of $10^{13}$ M$_\odot$ \cite{rhs99}. 
Of course, the presence of a massive central galaxy and its
effect on the distribution of neutrinos has not been included here,
but it may well be that 
an additional undetected baryonic component in cooling flow
clusters-- perhaps resulting from the cooling flow itself-- is
required.  I have also not considered the possibility
that there is at least one massive sterile neutrino-- a possibility
with some theoretical and experimental motivation \cite{bggm}.  
In any case, the
greatest part of the discrepancy in clusters would be accounted for by the
three types of active neutrinos if their mass is near 2 eV.

We will not have to
wait long for this possibility to be falsified (or confirmed).  Currently
planned $\beta$ decay experiments \cite{oskat} will push the
upper limit on the electron neutrino mass to a few tenths of an
electron volt within a few years.  If it turns out to be
the case that $m_\nu \approx 2$ eV, then, with MOND,
the old problem of clusters is solved.

\section*{Acknowledgements}

I am grateful to Moti Milgrom for helpful comments on the manuscript and
to Nasser Mohamed Ahmed for useful discussions on cooling flow clusters
and relevant numerical calculations.


\begin{thebibliography}{}

\bibitem [Aguirre, Schaye \& Quataert 2001] {asq} Aguirre A., Schaye J.,
   Quataert E., 2001, ApJ, 561, 520
\bibitem [Angus et al. 2007] {angeal07}Angus G.W., Shan, H.Y., 
   Zhao, H.S.; Famaey, B., 2007, ApJ, 654, L13
\bibitem [Arnaud \& Evrard 1999] {ae99} Arnaud M., Evrard A.E., 1999,
   MNRAS, 305, 631
\bibitem [Bilenky et al. 2003] {bggm} Bilenky S.M., Giunti C.,
   Grifols J.A., Mass\'o E., 2003, Phys.Reps., 379, 69 (hep-ph/0211462)
\bibitem [Clowe et al. 2006] {cleal06} Clowe D., 
  Bradac M.; Gonzalez A.H., Markevitch, M., Randall, S.W., Jones, C.; 
  Zaritsky, D.,2006, ApJ, 648, L109
\bibitem [David et al. 1990] {deal90} David L.P., Arnaud M.,
   Forman W., Jones C., 1990, ApJ, 356, 32
\bibitem [De Grandi \& Molendi 2002] {dgm} De Grandi S., Molendi S.,
   2002, ApJ, 567, 163
\bibitem [Edge \& Stewart 1991] {es91} Edge A., Stewart G.C., 1991,
   MNRAS, 252, 428
\bibitem [Fukuda et al. 1998] {fuku} Fukuda Y. et al., 1998, Phys.Rev.Lett.,
   81, 1562 
\bibitem [Gerbal et al. 1992] {geal92} Gerbal D., Durret F.,
   Lachi\`eze-Rey M., Lima-Neto G., 1992, A\&A, 262,395
\bibitem [Groom et al. 2000] {groom} Groom D.E. et al., 2000,
   Eur.Phys.J., C15, 1
\bibitem [Hoekstra, Yee \& Gladders 2004]{hoekeal} 
   Hoekstra H., Yee H.C., Gladders M.D, 2004, ApJ, 606, 67 
\bibitem [Ikebe et al. 2002] {ikeal} Ikebe Y., Reiprich TH.,
   B\"ohringer H., Tanaka Y., Kitayama T., 2002, A\&A, 383,773
\bibitem [Kofman et al. 1996] {keal} Kofman L, Klypin A., Pogosyan D.,
   Henry J.P., 1996, ApJ, 470, 102 
\bibitem [Kull, Treumann \& B\"ohringer 1996] {ktb96} Kull A., Treumann R.A.,
   B\"ohringer H., 1996, ApJ, 466, L1
\bibitem [Kull, Treumann \& B\"ohringer 1997] {ktb97} Kull A.,
   Treumann R.A., B\"ohringer H., 1997, ApJ, 484, 58
\bibitem [Landau \& Lifshitz 1980] {ll80} Landau L., Lifshitz E.M., 1980,
   Statistical Physics, 3rd Ed., Part 1, Elsevier Butterworth-Heinemann
   Oxford
\bibitem [Lloyd-Davies, Ponman \& Cannon 2000] {ldpc} Lloyd-Davies
  E.J., Ponman T.J., Cannon D.B., 2000, MNRAS, 315, 689
\bibitem [Lynden-Bell 1967] {dlb67} Lynden-Bell D., 1967, MNRAS, 136,101
\bibitem [Madsen \& Epstein 1984] {madep} Madsen J., Epstein, R.I., 1984,
   ApJ, 282, 11
\bibitem [McCarthy, Babul \& Balogh 2002] {mbb02} McCarthy I.G.,
   Babul A., Balogh M.L., 2002, ApJ
\bibitem [McGaugh 2004] {ssmcmb} McGaugh S.S., 2004, ApJ, 611, 26
\bibitem [Milgrom 1983] {m83} Milgrom, M., 1983, ApJ, 270, 365
\bibitem [Milgrom 1984] {m84} Milgrom, M., 1984, ApJ, 287, 571
\bibitem [Milgrom 1998] {m98} Milgrom, M., 1998, ApJ, 496, L89
\bibitem [Milgrom \& Sanders 2007] {ms07} Milgrom M., Sanders R.H., 2007
   ApJ, 
\bibitem [Mohr, Mathiesen \& Evrard 1999] {mme99} Mohr J.J., Mathiesen B.,
   Evrard A.E., 1999, ApJ, 517, 627
\bibitem [Mohr et al. 2000] {meal00} Mohr J.J., Reese E.D., Ellingson E.,
   Lewis A.D., Evrard A.E., 2000, ApJ, 544, 109
\bibitem [Nakajima \& Morikawa 2007] {nakmor} Nakajima M., Morikawa M.,
   2007, ApJ, 655, 135
\bibitem [Osipowicz et al. 2001] {oskat} Osipowicz A., et al. (KATRIN
   collaboration), hep-ex/0109033
\bibitem[Piffaretti et al. 2005]{pjkt} Piffaretti R., Jetzer Ph.,
   Kaastra J.S., Tamura T., 2005, A\&A, 433, 101
\bibitem [Ponmon, Cannon \& Navarro 1999] {pcn99} Ponmon T.J., Cannon D.B.,
   Navarro J.F., 1999, Nature, 397 135
\bibitem [Pontecouteau \& Silk 2005] {posi05} Pontecouteau E.,
   Silk J., 2005, MNRAS, 342, 901
\bibitem [Pratt et al. 2006] {preal} Pratt G.W., B\"ohringer H.,
   Croston J.H., Arnaud M., Borgani S., Finoguenov A., Temple R.F.,
   2007, A\&A, 461, 71
\bibitem [Reiprich \& B\"ohringer 2002] {rb02} Reiprich T.H., B\"ohringer H.,
   2002, ApJ, 567, 716
\bibitem [Sanders 1994]{rhs94} Sanders R.H., 1994, A\&A, 284, L31
\bibitem [Sanders 1999]{rhs99} Sanders R.H. 1999, ApJ, 512, L23-L26
\bibitem [Sanders 2003]{rhs03} Sanders R.H. 2003, MNRAS, 342, 901
\bibitem [Sanders \& McGaugh 2002]{sm02}Sanders R.H. \& McGaugh S.S.,
   2002, Ann.Rev.A\&A, 40, 263
\bibitem [Sanders \& Noordermeer 2007] {sn07} Sanders R.H., Noordermeer E.,
   2007, submitted MNRAS
\bibitem [Spergel et al. 2006] {seal06} Spergel D. et al 2006, ApJ
   (in press), astro-ph/0603449
\bibitem [The \& White 1988] {tw88} The L.S., White S.D.M., 1988, AJ, 95, 1642
\bibitem [Tremaine \& Gunn 1979] {tg79} Tremaine S.C., Gunn J.E., 1979,
   Phys.Rev.Lett., 42, 408
\bibitem [Treumann, Kull \& B\"ohringer 2000] {tkb} Treumann R.A., Kull A.,
   B\"ohringer H., 2000, New J. of Phys., 2, 11.1
\bibitem [Vikhlinin et al. 2005] {vikeal} Vikhlinin A.,Kravtsov A., 
   Forman W., Jones C., Markevitch M., Murray S.S., Van Speybroeck L.,
   2006, ApJ, 640, 691 
\bibitem [Zhao \& Famaey 2006] {zf06} Zhao H.S., Famaey, B., 2006,
   ApJ, 638,L9

\end{thebibliography}
\end{document}